\documentclass[twocolumn, times, tighten]{aastex62}
\usepackage{graphicx}
\usepackage{amsmath}
\usepackage[nopatch]{microtype}
\usepackage{booktabs}
\usepackage{xcolor}
\usepackage{tabularx}
\usepackage{array}
\usepackage{rotating}
\usepackage{makecell}
\usepackage{tablefootnote}
\usepackage{booktabs}
\usepackage{threeparttable}
\usepackage{soul}
\usepackage{lineno}
\usepackage{hyperref}
\usepackage[title]{appendix}

\def\apjs{ApJS}

\def\apjl{ApJL}

\def\aap{A\&A}

\begin{document}

\submitjournal{ApJ}
\shorttitle{} \shortauthors{Patra et al.}

\title{Evidence for {\it in-situ} acceleration of relativistic particles in the wings of X-shaped radio galaxies}

\correspondingauthor{Dusmanta Patra}
\author{Dusmanta Patra}
\affil{S N Bose National Centre for Basic Sciences, Kolkata, \it{700106}; India}
\email{dusmanta.phy@gmail.com}

\author{Gopal-Krishna}
\affil{UM-DAE Centre for Excellence in Basic Sciences, Mumbai, \it{400098}; India}

\author{Ravi Joshi}
\affil{Indian Institute of Astrophysics (IIA), Koramangala, Bangalore, \it{560034}; India}


\begin{abstract}
We report evidence for {\it in-situ} acceleration/re-acceleration of relativistic particles in 11 radio wings out of a total of 68 wings sufficiently well-resolved for spectral mapping, which belong to our sample of 40 X-shaped radio galaxies (XRGs). This representative XRG sample includes 15 XRGs newly reported here, which we selected from the LOTSS-DR2 survey, following well-defined criteria. 
The evidence for {\it in-situ} particle acceleration comes from the observed cessation of steepening, or even flattening (i.e., gradient reversal) of the spectral index profile along the lobe into the associated wing, as determined here by combining the LoTSS-DR2 (144 MHz) and FIRST (1.4 GHz) maps. Interestingly, the afore-mentioned trends in spectral gradient, indicative of {\it in-situ} particle acceleration, are mostly found to set in near the region where the lobe plasma stream bends to connect to the wing. Such a spatial coincidence with bending of radio lobe/tail has been noticed in recent years for just a couple of radio galaxies. The large increase in such examples, as reported here, is expected to give a fillip to observational, theoretical and numerical simulation follow-up investigations of this important clue about the occurrence of {\it in situ} particle acceleration in lobes of radio galaxies.
\end{abstract}


%

\section{Introduction}
\label{sec:intro}

In powerful radio galaxies the pair of radio lobes is energised by two nearly anti-parallel jets of comparable power, launched from the active nucleus in the core of the parent galaxy. Even when the jets undergo directional changes due to precession \citep[e.g.,][]{Krause2019MNRAS.482..240K, Britzen2023ApJ...951..106B}, the two lobes represent the long-term average of the directions of the jets feeding them. Thus, the two lobes usually exhibit a broadly symmetric disposition about the nucleus and together define the primary (i.e. the main) radio axis, roughly along which they are normally found to extend, unless sometimes swept aside by a strong external cross-wind. In a small fraction of radio galaxies, however, the two     
lobes are seen to bend away from the radio axis broadly in {\it opposite} directions, giving the appearance of a second pair of radio lobes which are fainter, 
diffuse and distinctly misaligned from the main radio axis defined by the two primary lobes. The pair of fainter radio lobes representing the off-axis emission is usually 
termed `secondary lobes' or `wings'. Such sources, exhibiting two mis-aligned pairs of radio lobes are known as X-shaped radio galaxies (XRGs) and they constitute 
$\sim  10\%$ of the radio galaxy population \citep[][]{Yang2019ApJS..245...17Y, Cheung2007AJ....133.2097C, Bera2020ApJS..251....9B}.
One possibility is that the secondary lobe pair, i.e. the two `wings’ form due to buoyancy-driven diversion of the back-flowing synchrotron plasma streaming in the 
primary lobes, as the two plasma streams impinge asymmetrically on the interstellar medium (ISM) of the parent elliptical galaxy \citep[e.g.,][]{Leahy1984MNRAS.210..929L, Leahy1992ersf.meet..307L, Worrall1995ApJ...449...93W, Capetti2002A&A...394...39C, Hodges-Kluck2010ApJ...710.1205H, Hodges2011ApJ...733...58H, Cotton2020MNRAS.495.1271C, Gopal-Krishna-Dhabade_2022A&A...663L...8G}.
However, this simple scenario may run into tension with the existence of some XRGs in which wings are longer than the primary lobes. A few prominent such examples are the radio galaxies: B2 0828+32 \citep{Parma1985A&AS...59..511P, Klein1995A&A...303..427K}, J1130+0058 \citep{Lal2019AJ....157..195L}, 3C 403 \citep{Black1992MNRAS.256..186B, Condon1995AJ....109.2318C, Dennett-Thorpe1999MNRAS.304..271D} and J2011-298 \citep{Bruno2024A&A...690A.160B}. 
Here it may also be relevant to recall the recent discovery of huge radio `chimneys' emanating almost orthogonally from the lobes of standard radio galaxies, namely J022830+382108 and J093032+311215 \citep{Gopal-Krishna-2024arXiv240311290G}.
The other popular model of XRGs, the `spin-flip' model \citep{Rottmann2001PhDT.......173R, Zier2001A&A...377...23Z, Dennett2002MNRAS.330..609D, Gergely2009ApJ...697.1621G, Lalakos2022ApJ...936L...5L} invokes a minor merger (or, accretion-disk instability leading to inhomogeneous mass accretion) and is particularly explored in \citep{Merritt2002Sci...297.1310M}. This model is known to encounter difficulty in accounting for the observed radio-optical alignments and the high ellipticity of the host galaxy in XRGs \citep[e.g.][]{Gopal2012RAA....12..127G, Bruno2024A&A...690A.160B}.
Another difficulty which lies in explaining the lateral offset between the two wings, as observed distinctly in several XRGs, can be overcome in a variant of the basic spin-flip model. This modified scenario, which seeks to reconcile this particular observational challenge with the basic spin-flip model, posits that the twin-jets prior to the spin-flip 
are diverted in opposite directions due to their dynamical interaction with the ISM of the host galaxy, which is already set in rotation, e.g., due to inspiralling of a dwarf galaxy \citep{Gopal2003ApJ...594L.103G, Gopal2012RAA....12..127G}.
Here, we also recall another model of XRGs which seeks to explain the wing formation as a consequence of the twin-jets interacting with the gaseous shells rotating around the host elliptical (\citet{Gopal1983Natur.303..217G, Gopal1984A&A...141...61G, Gopal2010ApJ...720L.155G, Gopal2012RAA....12..127G}; see also, \citet{Hardcastle2006MNRAS.370.1893H} for further evidence of jet-shell interaction).
As pointed out by \citet{Joshi2019ApJ...887..266J}, this scenario has the potential to explain not only the wings that sometimes grow longer than the associated primary lobes (see above), but also the intriguing finding that so often the wings are seen to start off rather close behind the tip of the primary lobe, far away from the host galaxy \citep{Saripalli2018ApJ...852...48S}. Of course, depending on the angular resolution of the map, such XRGs could even be classified as Z-shaped radio galaxies, 
ZRGs, which are conventionally explained in terms of jet precession (\citet{Riley1972MNRAS.157..349R, Ekers1978Natur.276..588E, Bridle1979ApJ...228L...9B, Miley1980ARA&A..18..165M, Zier2005MNRAS.364..583Z}; see also, \citet{Parma1985A&AS...59..511P}, for possible role of conical jet precession in the case of XRG). 
However, as seen below, pinning down the exact physical process responsible for the wing formation in XRGs is not critical for the main theme pursued in this paper and 
indeed more than one mechanisms may lead to the X-shaped radio morphology \citep[see, e.g.][]{Saripalli2018ApJ...852...48S, Lal2019AJ....157..195L, 
Parma1985A&AS...59..511P, Bruno2024A&A...690A.160B}. 
Adequate for the present purpose is just the observed fact that in XRGs, the two primary lobes are seen to bend in roughly opposite directions, connecting to their respective wings.

In all these XRG models, the radiating synchrotron plasma in the wings is expected to be older compared to that in the associated primary lobes and consequently one would expect to observe in the wings signs of synchrotron aging, i.e., a steeper radio spectrum. 
Yet in a sample of 26 XRGs, selected regions within the wings and in their primary lobes were found to be statistically indistinguishable in terms of radio spectral index (averaged over the individual regions) and cases of wings having spectra flatter than their primary lobes were inferred to be as common as the obverse \citep{Lal2019AJ....157..195L} \footnote{Curiously, these authors found the spectra of the wings to be systematically flatter compared to the spectra of the diffuse emission (of comparable surface brightness) associated with classical double radio sources.}. This statistics led these authors to conclude that in XRGs the pairs of primary lobes and wings exhibit no systematic spectral difference attributable to synchrotron ageing and hence the two pairs are being independently energised quasi-simultaneously by two twin-jets launched from a binary of supermassive black holes situated within the core of the parent galaxy \citep{Lal2007MNRAS.374.1085L, Lal2019AJ....157..195L}.
This interesting proposal, however, falls short of explaining the well known result, namely, the total lack of jets and/or terminal hot spots in the wings, even though they are frequently observed in the primary lobes. Here it is also relevant to note that in their work, spectral index was measured over a narrow frequency range (240 - 610 MHz) confined to the metre-waveband where plasma ageing due to synchrotron losses is anyway expected to be mild \citep[e.g.][]{Machalski2007A&A...462...43M, Hardcastle2013MNRAS.433.3364H, Bruno2024A&A...690A.160B}.
Nonetheless, that study underscored the need for more accurate and comprehensive spectral imaging studies of XRGs, as was recently performed by \citet[][hereafter Paper I]{Patra2023MNRAS.524.3270P}.
Independently, an extreme, counter-intuitive behaviour was found for the XRG 3C 223.1, which remains the most robust counter-example to the generic prediction of wings to have a steeper spectrum in comparison to their primary lobes. First noted by \citet{Rottmann2001PhDT.......173R}, this abnormality of 3C 223.1 has recently been robustly established using the high-resolution imaging of this XRG with VLA (5 and 8 GHz) and LOFAR (147 MHz), yielding an order-of-magnitude improvement in sensitivity, angular resolution and the frequency range covered. In particular, these observations revealed that in the wings radio spectrum becomes even flatter than that at the hot spots in the associated primary lobes \citep{Gopal-Krishna-Dhabade_2022A&A...663L...8G}. 
Subject to the caveats mentioned above (i.e., the absence of hot spot or jet in every wings known so far), this rare and counter-intuitive spectral pattern would be consistent with the proposal that in XRGs wings and primary lobes are powered independently by two active SMBH within the galactic core \citep{Lal2007MNRAS.374.1085L, Lal2019AJ....157..195L}.
Alternatively, this spectral behavior could be a manifestation of {\it in-situ} particle re-acceleration occurring in the wings, e.g., at the shocks in the wings, as recently seen in the relativistic magneto-hydrodynamic simulations of XRGs \citep[e.g.,][]{Giri2022A&A...662A...5G}. 
Therefore, our main motivation in Paper I was to find some more examples of this extreme phenomenon in XRGs. In that work we presented plots of spectral index ($\alpha^{1400}_{144}$) along 40 ridge-lines delineating the radio lobes in a well-defined sample of 25 XRGs. These ridge-lines extending from the outer edge of the primary lobe into the associated wing revealed at most one case (the West lobe of J1015+5944) consistent with the spectrum becoming flatter in the wing, compared to that near the hot spot of the primary lobe associated with the wing (the other potential case, namely the North lobe of  J1548+4451 was deemed uncertain since the hot spot in the primary lobe is not readily identifiable). 
Thus, in Paper I, it was concluded that XRGs having wings in which radio spectrum becomes flatter compared to the primary lobes, like the XRG 3C 223.1, are exceedingly rare. Nonetheless, the spectral index profiles measured along the 40 ridge-lines did reveal 5 ($\sim$ $10 - 15\%$) cases where some spectral flattening was seen towards the wing (see Sect. 4 below). 
A key motivation behind the present study is to verify that finding using an independent sample of XRGs and, furthermore, to continue searching for additional example(s) of anomalous spectral pattern of the kind observed in the XRG 3C 223.1. 
A related motivation was to look for observational clues on possible trigger(s) for the {\it in-situ} particle re-acceleration, signatures of which had been noticed in some XRG wings, as mentioned above (see also Sect. 4 \& 5 ).

Since lobe bending is an essential structural connect between a wing and its primary lobe, XRGs are the objects of choice for examining the role of lobe bending in triggering {\it in-situ} particle re-acceleration, as inferred in some theoretical studies and numerical simulations mentioned above. We report here the spectral imaging of a new unbiased sample of 15 XRGs (Sect. 2). Notes on the spectral imaging procedure are provided in Sect. 3 and the results obtained are presented in Sect. 4, followed by a brief discussion in Sect. 5 and the main conclusions in Sect. 6. The spectral index maps for the entire sample can be found in Appendix I.

\section{A new sample of 15 XRGs}
\label{sec:sample}

For extracting the XRG sample from the 144 MHz LoTSS-DR2 \citep{Shimwell2022A&A...659A...1S}, the first 3 filters applied were: (i) the catalogue values of the major and minor axes should exceed $24^{\prime\prime}$ and $10^{\prime\prime}$, respectively; (ii) the published dynamic range for that field should exceed $\sim$ 40:1 (both filters are to facilitate the wings' discernibility, see, also, \citet{Cheung2007AJ....133.2097C}) and (iii) the flux density is $>$ 500 mJy at 144 MHz. This led to a basic sample of 2428 sources whose structures were then inspected by us visually for the presence of a X- or Z-shaped appearance and the search yielded 123 such candidates. We next visually inspected their 1.4 GHz VLA FIRST survey maps\footnote {\href{http://sundog.stsci.edu}{http://sundog.stsci.edu}} \citep{Becker_first_1995ApJ...450..559B} and the 3.0 GHz VLASS maps\footnote {\href{https://science.nrao.edu/vlass}{https://science.nrao.edu/vlass}} (\citep{Lacy-vlass-2020PASP..132c5001L}), along with the LoTSS-DR2 maps
\footnote  {\href{https://lofar-surveys.org/dr2_release.html}{https://lofar-surveys.org/dr2\_release.html}} \citep{Shimwell2022A&A...659A...1S}. On examining the morphologies of all these candidates in the three major radio surveys, we short-listed 55 XRGs for which at least one wing is sufficiently well resolved to delineate its ridge-line. As explained in the next paragraph, this sample got further reduces to 15 XRGs upon imposing the condition that any flux missing in the 1.4 GHz FIRST map is an insignificant fraction of the total flux at that frequency.

\begin{table}   
\centering
\scriptsize
\addtolength{\tabcolsep}{-0.4em}
\caption{The present sample of 15 XRGs selected from LOTSS-DR2. }
\label{source-list}
\begin{tabular}{l c c l r r }
\hline
Name   of    &    RA       &  Dec        &   Redshift($z$)  & \multicolumn{2}{c}{Flux (mJy)  at 1.4 GHz}        \\
 the XRG          &   (J2000)   &  (J2000)    &         &    NVSS~~~         &  FIRST     \\
\hline           
J0845+4031 & 08 45 08.38    & +40 31 15.8   &   $0.429$   &   163.6 $\pm$ 5.5 &    154.8  \\
J1005+5523 & 10 05 17.68    & +55 23 37.3   &   $0.445 \pm 0.447 ^{\dagger}$  &   185.0 $\pm$ 6.5  &   177.2   \\
J1018+2914 & 10 18 27.29    & +29 14 18.5   &   0.389   &   444.3 $\pm$ 18.9  &   415.4    \\
J1027+4834 & 10 27 14.27    & +48 34 33.6   &   $0.503 \pm 0.322 ^{\dagger}$  &    56.2 $\pm$   2.0 &     55.0  \\
J1114+4133 & 11 14 25.82    & +41 33 11.5   &   $0.610 \pm 0.076 ^{\dagger}$  &   211.5 $\pm$   7.2 &    200.8   \\
J1134+3046 & 11 34 19.26    & +30 46 38.5   &   $0.246 \pm 0.0004$            &   224.4 $\pm$   8.5 &    264.7    \\
J1139+5745 & 11 39 29.49    & +57 45 20.9   &   $1.080 \pm 0.431 ^{\dagger}$  &    88.6 $\pm$   3.0 &     ~87.8    \\
J1214+4540 & 12 14 15.04    & +45 40 03.9   &   $0.796 \pm 0.103 ^{\dagger}$  &    92.7 $\pm$  3.1 &    90.5   \\ 
J1342+2547 & 13 42 45.33    & +25 47 11.6   &   $0.597 \pm 0.130 ^{\dagger}$  &   363.4$\pm$ 15.8 &  339.5      \\
J1348+4411 & 13 48 04.60    & +44 11 24.2   &   $0.26635 \pm 0.00004 $         &  157.7 $\pm$ 5.3 &    161.6    \\
J1354+4657 & 13 54 36.23    & +46 57 04.2   &   $0.180 \pm 0.034 ^{\dagger}$   &   99.8 $\pm$  3.4 &    97.1    \\
J1427+3247 & 14 27 58.73    & +32 47 41.5   &   $0.56962 \pm 0.00003 $         &  137.2 $\pm$9.8  &   123.8  \\
J1544+3044 & 15 44 13.33    & +30 44 01.8   &   $0.777 \pm 0.058 ^{\dagger}$   &  238.6 $\pm$ 8.6 &    224.8     \\ 
J1606+4517 & 16 06 38.87    & +45 17 37.1   &   $0.556 \pm 0.00004 $           &  115.6 $\pm$ 4.0 &    116.5  \\
J1613+3907 & 16 13 43.00    & +39 07 32.7   &   $0.97542 \pm 0.00070$          &  435.2 $\pm$  15.9 &    412.5   \\

\hline
\end{tabular}
\vspace{0.2cm}
${\dagger}$--photometric redshift.
\end{table}

The very similar beams of the 1.4 GHz FIRST ($\sim 5^{\prime\prime}$ ) and the 144 MHz LoTSS-DR2 ($6^{\prime\prime}$) surveys make them well-suited for spectral mapping. But, whereas the LoTSS observations have a uv spacing lower limit of $\sim 50\lambda$, ensuring a high sensitivity to any expected large-scale emission in these sources, the much larger values of the lower uv limit for the FIRST survey at 1.4 GHz means that the sensitivity may decline significantly for structures more extended than about 1 arcmin \citep{Becker_first_1995ApJ...450..559B}.
We have minimised the potential impact of such a possibility of `missing flux' by checking for compatibility of the FIRST (1.4 GHz) flux density with that measured in the NVSS (1.4 GHz) survey which has essentially full sensitivity for scales upto $\sim 10$ arcmin. Accordingly, we have only selected those XRGs whose quoted 1.4 GHz FIRST flux density either falls within 1.5$\sigma$ error of, or exceeds, that reported in the  1.4 GHz NVSS observations made with a much larger beam of 45 arcsec \citep{Condon_nvss_1998AJ....115.1693C}. This flux compatibility requirement reduced our sample from the 55 to 15 XRG which are thus deemed here suitable for spectral imaging, by combining their LoTSS-DR2 (144 MHz) and FIRST (1.4 GHz) maps (see Table \ref{source-list}). Note that the rms errors in the FIRST images of our sample are between 125 to 175 $\mu$Jy beam$^{-1}$, while for the LoTSS-DR2 images, the range is from 160 to 483 $\mu$Jy beam$^{-1}$.  We emphasize that in the entire sample selection process, we have taken no cognisance of any pre-existing spectral information on the sources examined.

    \begin{figure*}
        \centering
        \includegraphics[width=0.9\linewidth, height=0.65\linewidth]{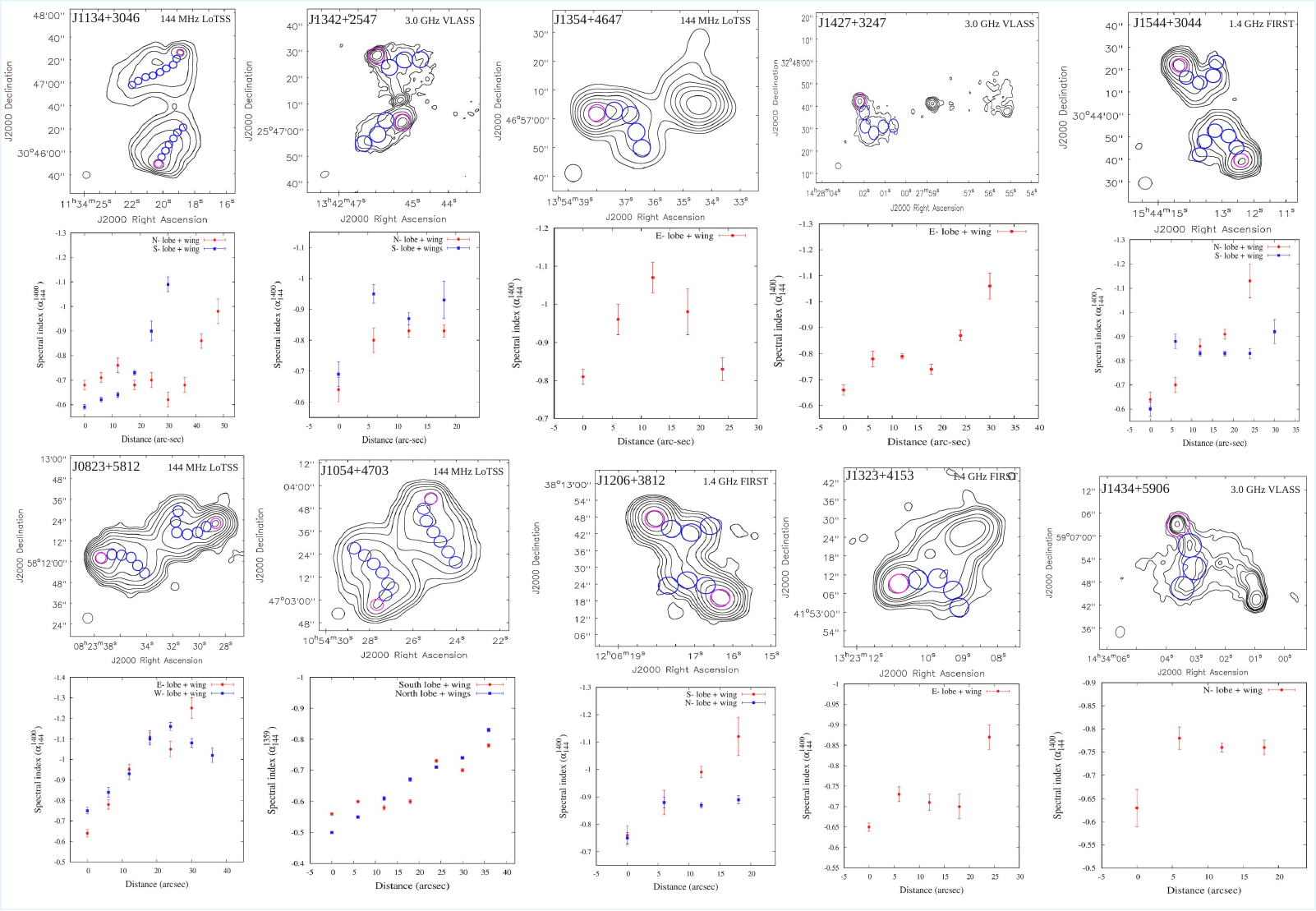}
        \caption{Radio maps showing the chain of contiguous circles of 6" diameter marking the ridge-line along each lobe and its associated wing. For each map, the spectral indices measured at the locations of the circles are shown in the panel directly below; the zero point on the horizontal axis corresponds to the position of the brightness peak in the primary lobe (hot spot), marked by a red circle in the lobe's map. The top two rows display the 5 XRGs from the present work and the lower two rows are for the 5 XRGs belonging to Paper I (see Sect. 4).}
        \label{fig:fig-1}
    \end{figure*}

\section{Data Analysis}
\label{sec:anlys}
Using the task CONVL in (AIPS)\footnote{\href{http://www.aips.nrao.edu/}{http://www.aips.nrao.edu/}}, we first smoothed 
the 1.4 GHz maps of the 15 XRGs to the $6"$ FWHM of their LoTSS-DR2 maps at 144 MHz. The geometry of the images at the two frequencies was then adjusted to match them in pixel sampling, coordinates and reference pixels, using the AIPS task HGEOM. 
Adopting the convention  $S \propto \nu^{\alpha}$, we then produced the spectral index ($\alpha$) map for each XRG by calculating $\alpha$ for each pixel in the two maps, using the AIPS task COMB. For this purpose, we have only considered the regions of the VLA/FIRST and LoTSS-DR2 maps within their 5$\sigma$ contours.

The rms error on $\alpha$ was computed using the relation:
\vskip 0.1cm
$\delta\alpha= \frac{1} {ln(\nu_1/\nu_2)} \sqrt{(\frac{\sigma S_1}{S_1})^2+(\frac{\sigma S_2}{S_2})^2}$ \\
\vskip 0.1cm
where $\sigma S$ is the rms error on flux density; $S_1$ and $S_2$ are the flux densities at frequency $\nu_1$ and $\nu_2$, respectively.

To display the spatial variation of $\alpha$, we drew a profile path (ridge-line) joining the brightest parts of the primary lobe and its associated wing. Fig. \ref{Fig-A} shows for all the 15 XRGs the spatial variation of $\alpha$, measured at between 3 to 9 circular regions along the extent of the profile path (marked as a chain of contiguous circles, each representing the $6^{\prime\prime}$ FWHM of the spectral index maps). Thus, $\alpha$ computed at each point along the profile path corresponds to the average over a $6^{\prime\prime}$ diameter circle.

\section{Results}
\label{sec:res}

For each of the 15 XRGs, the radio maps at the multiple frequencies are displayed in Appendix I (Fig. \ref{Fig-A}), in a single-row, 6-panel mosaic, 
together with the spectral index ($\alpha^{1400}_{144}$) and error maps, as well as the run of $\alpha^{1400}_{144}$ along the ridge-line(s) marked with contiguous circles in one of the displayed radio maps which is deemed best suited for delineating the wing(s).

The two main questions we pose here are: (i) how often are discernible signatures of {\it in-situ} particle acceleration present in the spectral index profiles of the wings, and (ii) does the particle acceleration get triggered at some preferred location(s) within the lobe-wing configuration?
As seen from Fig. \ref{Fig-A}, for 28 out of the total 30 lobes of the 15 XRGs in our sample, we could delineate with reasonable clarity the ridge-line extending from the (primary) lobe into the associated wing. 
It is worth mentioning that the spectral index profile along a ridge-line would provide a more robust expression of the spectral trend, because a ridge-line traces the brightest parts of the lobe-wing configuration. Any spectral steepening between a primary lobe and its associated wing (which would signify radiative ageing), or, alternatively, a cessation of the spectral steepening, or even a reversal of the spectral steepening from the lobe towards the wing (which would indicate {\it in-situ} acceleration/re-acceleration of the radiating particles) should be more readily discernible in such ridge-line profiles of $\alpha$ (Sect. \ref{sec:anlys}). 
From Fig. A1, most of the $\alpha$ profiles are broadly consistent with monotonous spectral steepening along the ridge-line from the brightness peak of the primary lobe (hot/warm spot) towards the associated wing. This is broadly in consonance with the leading theoretical models of XRGs
\citep{Gopal2012RAA....12..127G} (see Sect. 1). Interestingly, however, five of these 15 XRGs show signs of departure from this common trend. More specifically, in the present sample of total 28 ridge-lines along which $\alpha$ profile could be determined, such `anomalous' spectral gradient is detected in six ridge-lines as displayed in the upper half of Fig. \ref{fig:fig-1}, with comments given below.

{\bf J1134+3046:} The LoTSS-DR2  map of this XRG exhibits wing formation due to an eastward bending of the N-lobe. The bend occurs mainly in the region of the second and third circles, starting from the red circle marking the hot spot of the N-lobe (Fig. \ref{fig:fig-1}). The measured $\alpha$ values {\bf for} these circles show onset of spectral flattening and this trend continues for $\sim 20"$, whereafter the spectral steepening resumes (Fig. \ref{fig:fig-1}). In contrast, the southern wing is straight and its $\alpha$-profile shows a monotonous spectral steepening along the wing. It is worth pointing out that a comparison of the FIRST and NVSS flux densities of this XRG (Table 1) excludes any possibility of `missing flux' in the FIRST map which we have used in combination with the LoTSS-DR2 map, to determine the $\alpha$ profile displayed in Fig. \ref{fig:fig-1} (see Sect. 2).

{\bf J1342+2547:} Each wing of this XRG displays a fairly sharp bend $\sim 6''$ behind the hot spot and onset of spectral flattening is observed at that location in each lobe.

{\bf J1354+4657:} In the $\alpha$-profile of the E-wing, significant spectral flattening is seen to set in $\sim 12''$ behind the hot spot. This is also the location where the wing displays a sharp bend southward. 

{\bf J1427+3247:} The $\alpha$-profile along the E-wing shows a localised flattening $\sim$ $15''$ behind the hot spot where the radio structure is seen to undergo a conspicuous bend.  

{\bf J1544+3044:} Although both lobes exhibit a well-developed wing with a distinct bend, their $\alpha$-profiles are markedly different. Whereas a monotonous spectral steepening is observed along the N-wing, the $\alpha$-profile in the S-wing becomes flat after $\sim 12''$ from the hot spot where the ridge-line bends sharply. Additional comments on this spectral contrast are provided in Sect. 5. 

With the intent of augmenting the above set of 5 XRGs, we have revisited our recently published $\alpha^{1400}_{144}$ profiles of 40 ridge-lines associated with 25 XRGs (Paper I). This has yielded a set of another 5 XRGs whose ridge-line(s) exhibit fairly distinct signatures of {\it in-situ} particle acceleration, manifesting as cessation of spectral steepening (or, even a reversal of the spectral gradient), as one proceeds from the primary lobe towards the associated wing. The $\alpha$-profiles for the 5 ridge-lines belonging to these 5 XRGs,  namely J0823+5812 (W-lobe),  J1054+4703 (S-lobe), J1206+3812 (N-lobe), J1323+4153 (E-lobe) and J1434+5906 (N-lobe)), taken from Paper I, are displayed in the lower half of Fig. \ref{fig:fig-1}. Again, it is found that in essentially all these cases, the onset of the above-mentioned spectral signatures of {\it in situ} particle acceleration is spatially correlated with the apparent bending of the synchrotron plasma stream from the primary lobe towards the wing.

\section{Discussion}
As seen from Fig. \ref{Fig-A} and in general agreement with Paper I, we find in the present sample of 15 XRG, just one case (N-lobe of J1134+3046) for which $\alpha^{144}_{1400}$ in the wing becomes (marginally) flatter compared to that in the associated primary lobe. As mentioned in Sect. 1, such a counter-intuitive spectral pattern has been firmly established for the XRG 3C 223.1 \citep{Gopal-Krishna-Dhabade_2022A&A...663L...8G}. Secondly, the set of 10 XRGs shown in Fig. \ref{fig:fig-1} exhibits fairly robust spectral signatures of {\it in-situ} particle acceleration occurring in their wings, manifesting as cessation of spectral steepening, or even a localised reversal of spectral gradient along the ridge-line from the lobe towards the wing (Sect. 4). Since in our entire sample of 40 XRGs, such signatures are seen in 11 out of the total 68 ridge-lines for which $\alpha^{1400}_{144}$ profiles could be determined (Paper I and Fig. 1 in the present work), this sets a lower limit of $\sim 16\%$ for the fraction of wings in which {\it in-situ} particle acceleration is likely occurring. Remarkably, as mentioned above, the onset of spectral flattening (or, cessation of spectral steepening) is typically seen to coincide with the location of the bend in the ridge-line that runs through the primary lobe and the wing associated with it. 
 
It is pertinent to recall that very few examples of such spatial coincidence have been reported so far.
One prime example, already mentioned above, is the XRG 3C223.1 \citep[][]{Gopal-Krishna-Dhabade_2022A&A...663L...8G}. The other potentially similar case highlighted by these authors is the XRG PKS 2014-55 \citep {Cotton2020MNRAS.495.1271C}. The remaining two known cases are: the Wide-Angle-Tail (WAT) radio galaxy MRC 0600-399 (termed `double-scythe', \citet{Chibueze2021Natur.593...47C, Bagchi2006Sci...314..791B}) and the WAT `GReEt' in the Abell cluster 1033 \citep{de_Gasperin2017SciA....3E1634D, Edler2022A&A...666A...3E}. The case of GreET is particularly striking, since {\it both} major bends observed in its well-collimated northern radio tail are found to be accompanied by onset of radio spectral flattening \citep{Gopal-Krishna-Witta-2024JApA...45...12G}. 

It may be recalled that from an observational perspective, the issue of {\it in-situ} particle acceleration in the lobes of radio galaxies has been discussed for nearly 
5 decades \citep[][and references therein]{Gopal-Krishna-Witta-2024JApA...45...12G}. 
The emerging picture is that different physical mechanisms might dominate under different settings and corresponding radio morphologies. 
For instance, recent 3D-relativistic MHD simulations of evolution of the jet/cocoon structure in classical double radio sources have underscored the efficacy of shocks 
and turbulence as the dominant process for particle re-acceleration in the jets and the well-aligned lobes 
\citep[][]{Dubey2023ApJ...952....1D, Mukherjee2021MNRAS.505.2267M, Yates-Jones2022MNRAS.511.5225Y}.
On the other hand,  in the cluster radio galaxy MRC 0600-399 whose jet is seen to bend and form a highly unusual ``double scythe" radio structure, the spectral flattening observed near the bend has been interpreted in terms of {\it in-situ} particle re-acceleration of a different physical origin \citep{Chibueze2021Natur.593...47C}. 
The 3D MHD simulations by these authors show that magnetic reconnection leading to efficient {\it in-situ} particle re-acceleration occurs as the jet bends upon collision with the layer of magnetic field of the intra-cluster medium, compressed due to the jet's impact and thereby producing a current sheet. 
A more generic case of bending of lobe plasma stream is exemplified by the wing formation in XRGs. 
For this circumstance, shear acceleration \citep{Berezhko&Krymskii1981SvAL....7..352B, Rieger2002A&A...396..833R, Rieger2007Ap&SS.309..119R, Sironi2021ApJ...907L..44S} could be a naturally viable mechanism for initiating spectral flattening near the bend 
\citep[see, also][]{Giri2022A&A...662A...5G, Ostrowski2000MNRAS.312..579O}.
Therefore, detailed spectral imaging of XRGs could be an important stepping stone towards unravelling the working of this mechanism.

We would like to end by drawing attention to a rather intriguing asymmetry observed in two of the XRGs in our sample, namely J1544+3044 and J0823+5812. In each of these XRGs, although the two lobes appear similar in terms of bending, spectral flattening (near the bend) is observed in just one of them (Fig. 1). Thus, it seems that the evidence for triggering of particle re-acceleration near the bend is discernible in only one of the two lobes in these two XRGs. Without speculating on the origin of this spectral asymmetry between the twin lobes, which needs further investigation, we merely point out here that even cases of classical double (FR II) radio galaxies are known where the two lobes exhibit grossly different spectral index gradients, despite their similar structural appearances. Two such examples are the radio galaxies 3C 68.2 and 3C 322 \citep{Leahy1989MNRAS.239..401L}.

\section{Conclusions}
The work presented here represents a large expansion of the emerging observational evidence accrued during the recent years, for {\it in-situ} acceleration/ re-acceleration of relativistic particles in the lobes of radio galaxies, probably triggered due to bending of synchrotron plasma streaming in the lobes. Because in XRGs, bulk-bending of the synchrotron plasma stream is always present and seen to provide a spatial connect between the primary lobe and the wing, we have examined here its possible influence on radio spectral index gradient along the primary lobes and their associated wings, as observed in two representative samples of XRGs reported here and in Paper I. The present investigation of these two well-defined sets comprising 40 XRGs in all, has revealed the signatures of {\it in-situ} particle acceleration in 11 of their wings, manifesting as plateauing (even flattening) of the spectral index towards the wing. Remarkably, in practically all these 11 cases, such spectral transition is found to spatially coincide with the bending of the radio structure. This significant expansion of observational evidence for such a spatial correlation is expected to provide impetus for detailed studies of this important plasma process occurring in the lobes of radio galaxies.

\section{acknowledgement}
GK thanks Indian National Science Academy for a Senior Scientist position under which a part of this work was carried out. GK and DP acknowledge the Indian Institute of Astrophysics for the hospitality during their visit when this work was initiated.

\bibliography{references}{}
\bibliographystyle{aasjournal}


\appendix
\counterwithin{figure}{section}
\renewcommand\thefigure{A\arabic{figure}}
\section{I}
\label{sec:appndx}

\begin{figure*}
\begin{center}
\includegraphics[width=15 cm]{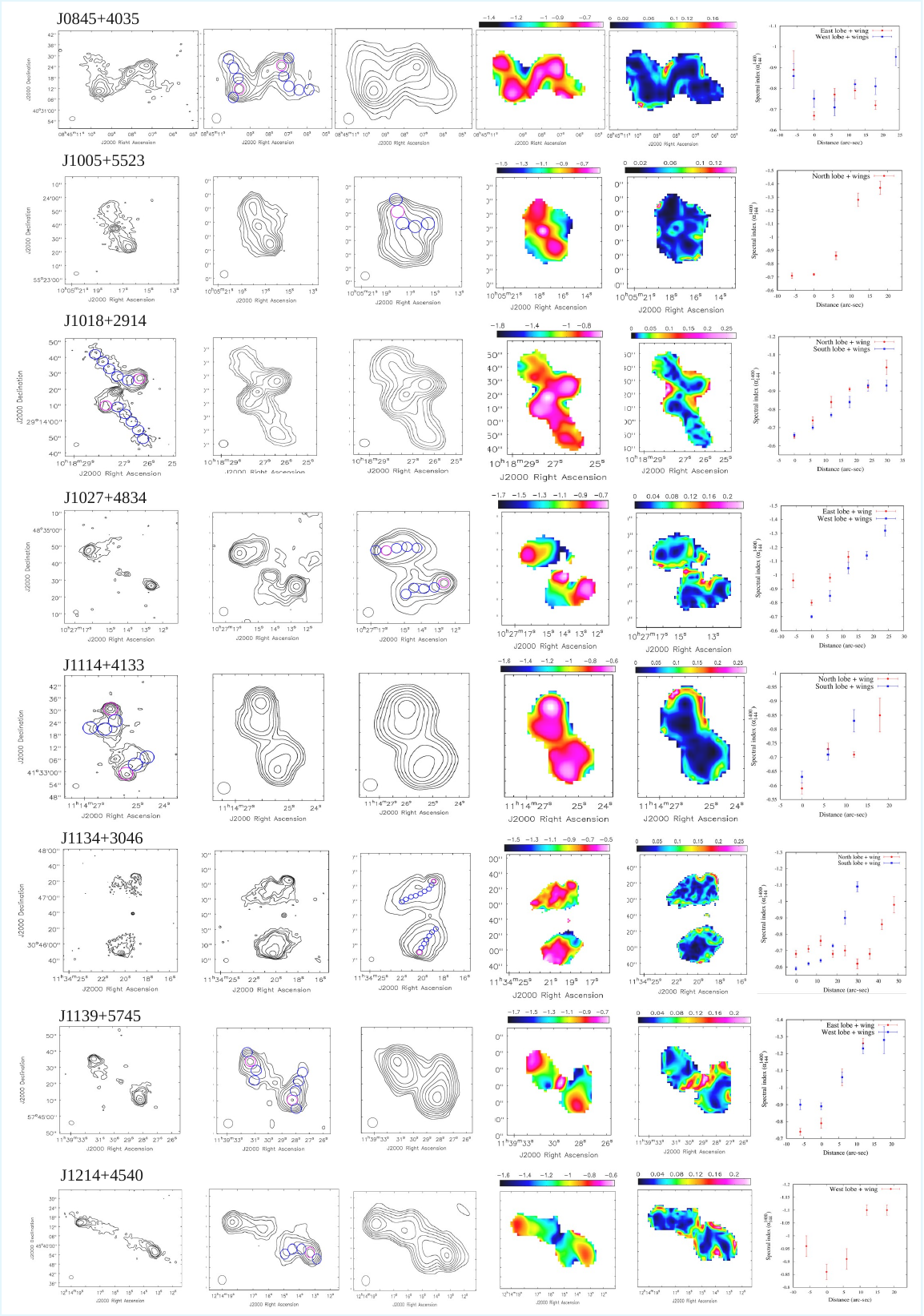}
\caption{The radio contour maps (VLASS, FIRST, and LoTSS-DR2), spectral-index (144 MHz -- 1.4 GHz) image, spectral-index error map, and the corresponding spectral index profile measured along the ridge-line (s) marked in the radio contour maps, are displayed row-wise for each XRG  in our sample of 15 XRGs. The name of the source is mentioned at the top of the first panel in each row. The last panel in each row displays spectral index at several points along the ridge-line(s). These values represent average taken over a circular regions of 6 arcsec diameter (the beamsize) as shown along the corresponding ridge-line displayed in one of the three radio maps deemed best suited for identifying the wing component.
} 
\label{Fig-A}
\end{center}
\end{figure*}
\addtocounter{figure}{-1}
\begin{figure*}
\begin{center}
\includegraphics[width=15 cm]{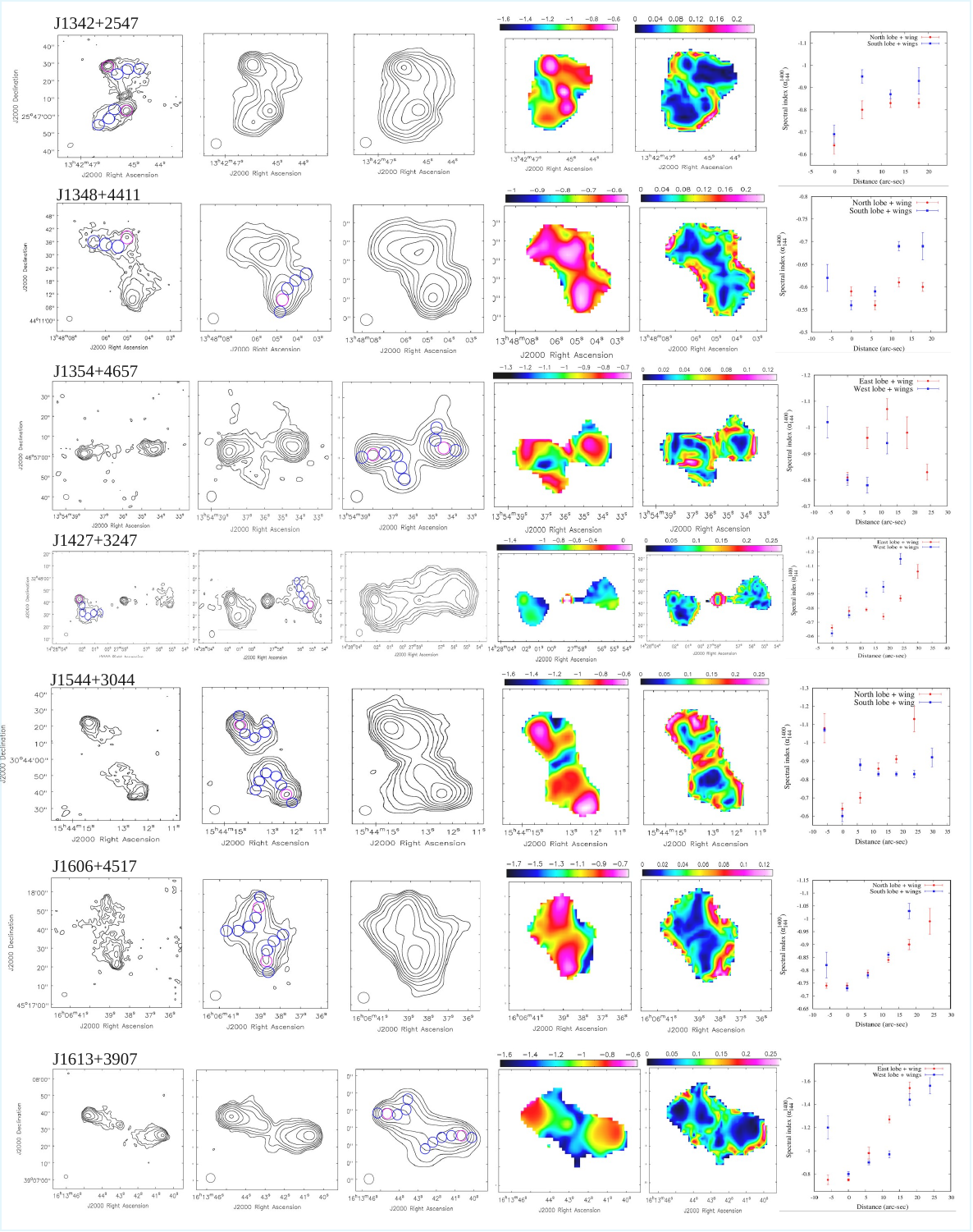}
\caption{Continued}  
\end{center}
\end{figure*}

\end{document}